# Enhancing search engine precision and user experience through sentiment-based polysemy resolution

Mike Nkongolo*

1Department of Informatics, University of Pretoria, Gauteng, South Africa

**Correspondence**

*Wa Nkongolo Mike Nkongolo, Faculty of Engineering, Built Environment and Information Technology

 Email: mike.wankongolo@up.ac.za

**Abstract**

In the era of information overload and the growing importance of personalized content delivery, users and businesses are constantly seeking ways to improve the accuracy and relevance of search results and content recommendations. Sentiment analysis and polysemy resolution are crucial in achieving this goal. With the proliferation of digital content and the need for efficient information retrieval, the study's insights can be applied to various domains, including news services, e-commerce, and digital marketing, to provide users with more meaningful and tailored experiences. The study addresses the common problem of polysemy in search engines, where the same keyword may have multiple meanings. It proposes a solution to this issue by embedding a smart search function into the search engine, which can differentiate between different meanings based on sentiment. The study leverages sentiment analysis, a powerful natural language processing (NLP) technique, to classify and categorize news articles based on their emotional tone. This can provide more insightful and nuanced search results. The article reports an impressive accuracy rate of 85% for the proposed smart search function, which outperforms conventional search engines. This indicates the effectiveness of the sentiment-based approach. The research explores multiple sentiment analysis models, including Sentistrength and Valence Aware Dictionary for Sentiment Reasoning (VADER), to determine the best-performing approach. The study uses real-world data from the British Broadcasting Corporation (BBC) news site, making the findings more applicable to real-world scenarios. The findings can be applied to enhance search engines, making them more capable of understanding the context and intent behind users' queries. This can lead to better search results that are more aligned with what users are looking for. The study's approach can be applied in the field of news classification and categorization. News agencies and online platforms can use sentiment analysis to automatically categorize news articles and improve the accuracy of their recommendations to users. The smart search function can improve the user experience by reducing the need to sift through irrelevant search results. This is particularly important in an age where information overload is common.

**KEYWORDS:** search engine optimization; sentiment technology; information retrieval; word2vec; NLTK; BM25; inverted index

# 1 | INTRODUCTION

Over the past few years, there have been significant advancements in natural language processing (NLP) [1]. These advancements have been driven by improvements in linguistic models that can predict words, characters, and sentences from textual data [1, 2]. Chatbot models are considered the most efficient NLP programs, demonstrating accurate performance across various existing datasets for different NLP problems, including question answering, translation, news articles generation, sentiment analysis (opinion mining), and unscrambling words. This has been highlighted in literature by several researchers [3, 4, 5]. Nevertheless, the current evaluation and training approach for NLP favors only algorithms that have access to large datasets. In addition, the polysemy of textual patterns can negatively impact the performance of deployed NLP models.

To achieve human-like language capability, an NLP program must employ complex and disruptive technologies, while also addressing the need for feature engineering [1, 2]. The study aimed to address the challenges in generating accurate news using NLP by bridging the gap between NLP expectations and the difficulties encountered. The main methodology used in NLP is the engineering of textual patterns. However, the study recognized the shortcomings of implementing standard NLP to textual data without feature engineering. To address this, the study proposed a methodology that incorporates feature engineering by utilizing NLP models like embeddings and tokenizers to extract relevant features from existing data.

The proposed methodology aims to improve computational efficiency by reducing empirical errors and increasing accuracy levels. This article advocates for the adoption of an NLP system that focuses on extracting relevant textual data from the web for sentiment analysis. Although NLP has achieved notable progress, it is not yet equipped to address real-world problems with accuracy [3, 6], since its success is dependent on big data, evaluation metrics, and training approaches that favor probabilistic and heuristics learning. There are currently numerous well-known news websites available [7, 8], and most of them include an integrated search feature. Two popular news sites, China (www.chinanews.com) and BBC (www.bbc.co.uk), have built-in search functions, but these are limited and do not include advanced options like sentiment analysis. This study aims to design and implement a smart search engine with sentiment analysis capabilities to determine the opinion of search results and categorize them as negative, positive, or neutral. Such a search engine can automatically extract brand visibility or reputation from the internet in real-time by scoring search mentions positively or negatively.

Thus, the implementation of a smart search engine that incorporates sentiment analysis can provide a real-time understanding of people's attitudes towards specific brands during their internet news search. The sentiment analysis results can provide relevant insights to improve a brand's market share, competitive advantage, and reputation. Additionally, search engines with smart functions and sentiment analysis can influence consumers to purchase products with a positive reputation.

In the context of news classification and categorization, a smart function based on sentiment analysis can foster brand trust and elevate a brand's reputation if the sentiment analysis results are positive. Text categorization and classification share many similarities, with the latter being a subset of the former [9, 10, 11]. In text categorization, the first step is to represent the text by preprocessing the documents and creating a vector space containing the words present in the documents using the bag of words (BoW) model.

Consequently, classification of text documents relies on the proximity of keyword vectors, with the significance of keywords in the documents often determined by weighting schemes such as term frequency or word frequency. In contrast, sentiment analysis involves identifying relevant keywords in a textual document using linguistic patterns [11].

The sentiment classification field has adopted the n-gram technique, which involves dividing sentences into tokens and using the sequence of tokens for text representation. Additionally, the part of speech (POS) technique is employed to tag words with their grammatical attributes, such as nouns, adverbs, verbs, or adjectives. These types of representations are commonly used in sentiment classification research [9, 11, 12]. This study employed a smart function to obtain unambiguous search results from the BBC search engine, and then conducted sentiment analysis to assess the reputation of BBC news.

Users typically use keywords to search for news and expect the most relevant results. However, search engines often prioritize results based on their popularity rather than relevance, leading to a multi-connotation issue where desired results cannot be obtained due to multiple meanings of the keywords. Search engines like Google, Baidu, and Yandex prioritize recent or credible results, but these may not always be relevant to the user's needs. For example, searching for information about "apples" may yield results about both fruits and phones. The issue described is related to the problem of polysemy [8], which refers to multiple meanings of a word.

This challenge can be addressed by enabling users to provide more specific information to refine their search. However, only a few studies have addressed this issue. Some studies have utilized techniques such as anomaly detection and neural networks to classify search results that were manually collected [13].

This study aimed to optimize the categorization and classification of search results by implementing a search function on the search engine, resulting in fast and automated data collection. The proposed method also integrated sentiment analysis to determine the sentimentality of BBC news. The goal was to present a computational pipeline capable of quickly finding the desired data on the internet to classify patterns and analyze an entity's web presence. To achieve this, a search function was employed to extract BBC news and classify it into three categories (Covid, Vaccine, and Travel), and sentiment analysis algorithms were used to detect the news polarity as either negative, neutral, or positive. By using sentiment analysis as a proxy, this research utilized a classified search engine to understand the polarity assigned to a web entity. A database with various categories of news was created, but the search function was not able to search the entire internet.

However, it worked together with the search engine to automatically gather data. Each news item was tagged in the database to make sentiment analysis computation easier on the labeled data. The tags were fixed and divided into different categories. This article proposes a framework for processing textual data that brings engineering practices and paradigms to NLP in order to extract sentiment from web news. The proposed search function is envisioned as an optimized algorithmic pipeline that provides the most relevant results from the search engine. This article discusses how data quality and quantity can be addressed in engineering textual patterns, and emphasizes the importance of including unstructured feature engineering, which is currently not widely available due to limited packages, in NLP problems [14].

## 1.1 | Research aim

The main goal of this study was to create a search function that could gather BBC news from the internet, while also minimizing polysemous results. This search function was integrated into the BBC search engine, and the collected data was stored in a database. Sentiment analysis algorithms were then applied to this database to determine the polarity of the BBC news. The performance of the sentiment analysis was evaluated using metrics such as precision, accuracy, F1 score, and recall. The following research questions were the focus of this study:

- What effect do smart functions have on a search engine for news categorization?

- Which sentiment analysis model is the most suitable to classify news?

- How does NLP pre-process news data?

The implementation of the smart function not only aims to improve the user's search experience, but also facilitates the automated collection and sentiment analysis of relevant BBC news data. By reducing the polysemous issues, the function enables quick and efficient collection of data, which can be further categorized through automated sentiment analysis. The results of the study demonstrate that VADER was the most accurate and precise sentiment analysis model suitable for integration into a search engine for automated sentiment analysis.

Moreover, this research highlights the importance of NLP techniques in the processing of unstructured data. By engineering unstructured patterns and standardizing file formats in the database, data preparation and analysis become more manageable. This NLP-based approach enables the utilization of native formats of unstructured patterns, which can be processed and analyzed efficiently to provide insights into sentimentality of BBC news data. Therefore, this study provides a comprehensive framework for the collection, processing, and sentiment analysis of unstructured data, which can be adapted and utilized for other similar applications.

## 1.2 | Research rationale and contribution

Polysemy is a phenomenon in language where a word can have multiple meanings or interpretations [15]. This can lead to ambiguity and confusion in NLP tasks like search queries, where the context of the query may not always be clear. Sentiment analysis of search results can help with this problem by providing additional context and clues to disambiguate the meaning of a query. Sentiment analysis involves analyzing the emotional tone or sentiment of a piece of text by using NLP techniques to classify it as positive, negative, or neutral [16]. When applied to search results, sentiment analysis can provide valuable information about the context and connotations of the words in the results. For example, consider a search query for the word *bak*.

Depending on the context, this could refer to a financial institution or the side of a river. Sentiment analysis of the search results could help determine which meaning is most likely based on the emotional tone of the text. If the majority of the results are associated with positive financial news or reviews, it is more likely that the user was searching for a financial institution. Conversely, if the results are associated with negative river pollution or natural disaster news, it is more likely that the user was searching for the side of a river. In summary, sentiment analysis can help with the problem of polysemy by providing additional context and emotional clues to disambiguate the meaning of a search query. By analyzing the sentiment of search results, it is possible to infer the user's intended meaning and provide more accurate and relevant results.

The aim of this research paper is threefold. Firstly, it aims to present a solution for the polysemous problems encountered in search engines. Secondly, it seeks to identify the requirements that optimized search functions must meet. Thirdly, it argues for the use of optimized search functions to retrieve relevant data and conduct sentiment analysis to evaluate the strength of an entity's web presence. The paper introduces a search function that can be utilized for search engine optimization in practical applications. The research contributes to the fields of opinion mining and search engine optimization by presenting a relevant theory that can be easily implemented and tested for news categorization and classification. The primary objective is to present a computational methodology that analyzes the web presence of an entity to study its sentimentality using sentiment analysis as a proxy. The paper introduces such a framework and evaluates its performance on BBC news for news categorization and classification. The design of the study is depicted in Figure 1. The article is organized into several sections, including the background (Section 2), research methodology (Section 3), and results (Section 4). The conclusion and recommendations are presented in Section 5.

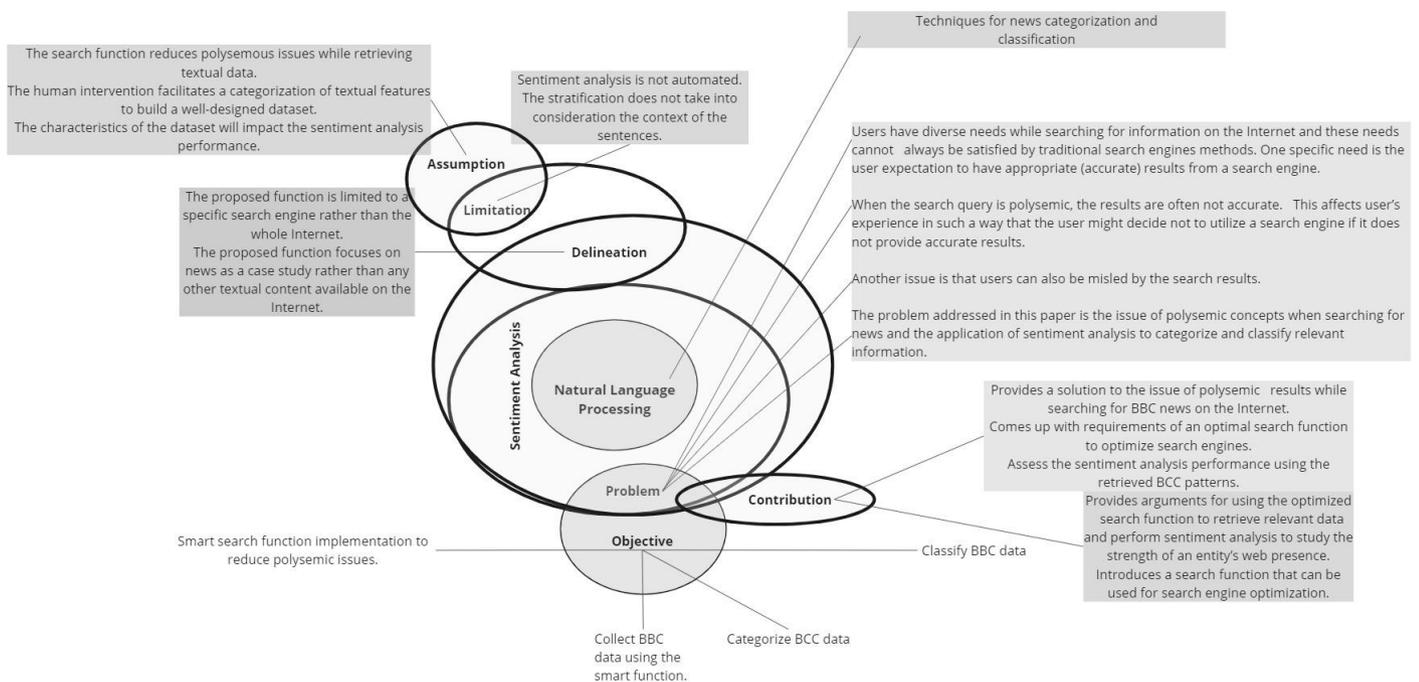

**FIGURE 1** The study design

## 2  | BACKGROUND

Although NLP has made significant advancements in achieving human-level performance in various tasks, researchers who seek to apply NLP to real-world problems still encounter challenges that demand more data annotation and training. To address this issue, an iterative and human-centered framework for NLP that incorporates feature engineering has been proposed.

The proposed framework, which adopts an NLP approach that focuses on the challenges of engineering textual data, holds promise for analyzing BBC news to determine their sentimentality. With this approach, it is expected that systems can be implemented to capture or understand the meaning of analyzed text [12], thus mitigating technical issues in the proposed computational framework [17].

One major obstacle in developing robust NLP systems is the "symbolic problem", in which symbols are interpreted based on their morphology rather than their intended meaning. This complication presents a significant challenge in assessing the capacity of computational machines to comprehend word meanings [17, 18].

NLP systems assume that the meaning of a sentence can be determined by analyzing a set of words in it. However, in technical and textual contexts, this assumption may not hold true since generalization is not always closely associated with sentence meaning. Furthermore, NLP systems do not learn effectively from training features and require significant retraining to improve the model's performance, as reported in recent studies [18, 19, 20].

The semantic disadvantage can be attributed to the evaluation metrics and current training methods that do not promote the implementation of human-like generalization models. As a result, NLP systems learn heuristically rather than learning the expected generalizations [19]. To address this issue, the proposed engineering approach involves human supervision and intervention, along with computational techniques such as annotation, tagging, and textual structuring, to facilitate sentiment analysis computation. Additionally, machine learning algorithms can be implemented as a tagging and forecasting approach to mitigate spurious correlations and heuristics that affect NLP learning.

In a study conducted by Yaxin Bi [11], a novel approach was introduced for sentiment analysis that utilized a belief functions framework to merge sentiment analysis outputs. The study evaluated the effectiveness of sentiment stratification individually and in combination. The results showed that the combined algorithms performed better than single classifiers across five review datasets. Yaxin Bi [11] developed a dichotomous ensemble learning method for sentiment analysis by using a triplet evidential scheme [9] to formulate negative and positive polarities, and negative and positive propositions for neutral polarity representation. This approach reduced the negative effects of neutral sentiments through ensemble learning and classified patterns with evidential reasoning schemes.

The proposed approach achieved a reduction in both the training time's complexity and cost, even when neutral polarity was not considered. The method represented text reviews using a BoW approach and performed ten-fold cross validation. The proposed methodology was evaluated using the F-score metric, and a total of 10,000 algorithms of triplet belief functions were utilized through three combination techniques with two groups of nine and eight learning classifiers. The results showed that the proposed method achieved an accuracy of 87%-90%, outperforming existing techniques [11]. Jalil et al [21] employed both deep learning and machine learning techniques to perform sentiment analysis on data related to the Corona Virus Disease2019 (COVID-19). Their experiments yielded an accuracy of 93% to 95%, which was higher than that of other techniques used in the study. In another study, Reddy et al [22] applied an NLP algorithm to segment a collection of medical publications. The methodology involved segmenting all the words and the associated information.

Pawade [23] demonstrated the benefits of a search engine optimization approach by using Google to extract essential information such as users' location and time spent on each website. Although the study did not involve pattern matching, it highlighted the potential of NLP algorithms in enhancing search engine optimization.

In the proposed methodology, user behavior analysis can be utilized to enhance the search engine experience by predicting user intent when searching for polysemous keywords. For example, if a user frequently searches for animal-related information, and inputs the keyword Jaguar, the search engine can predict that the user is likely searching for animal-related data, rather than information about the car brand. By analyzing user behavior and preferences, the search engine can improve its ability to provide relevant and accurate search results. This approach is like previous studies that have used user behavior analysis to enhance search engine optimization and improve the quality of user experience [24].

Yaxin Bi [10] conducted an experiment to evaluate the relationship between accuracy and diversity of classifiers using pairwise and non-pairwise diversity measures and evidential combination rules. The study utilized Yager's and Proportion's rules to generate ensemble learning negative classifiers, which did not prioritize minimizing the error of selected classifiers. Empirical results showed that increasing diversity decreased the accuracy of ensemble learning. However, the study did not investigate the behaviors of member classifiers concerning the efficiency of the ensembling scheme. Huang et al [25] introduced a novel approach, called the inverted index method, for efficient execution of temporal queries, which was demonstrated on a COVID-19 dataset [25]. The proposed method utilized a reverse sort-index approach, enabling real-time query processing to facilitate COVID-19 research.

The authors developed several categories of queries, including non-temporal, relative temporal, and absolute temporal queries. The results of the experiment suggest that the reverse sort indexing method is more efficient than current techniques in facilitating fast-time query execution for search engines. The inverted index is capable of rapidly retrieving non-polysemous big data. The proposed methodology is limited to searching and storing BBC news search results in the database. In the future, this approach can be applied to other search engines to extract more data. Nevertheless, sorting indexes, as highlighted by Huang et al [25], remains an essential aspect of search engine optimization.

Xu Z et al [20] devised a text-based approach for aircraft fault diagnosis by employing Word2vec and Convolutional Neural Networks (CNNs). The experiments utilized a large corpus of text files, and Word2vec was employed to retrieve textual pattern vectors which were then passed to the CNNs for final decision making. A Cloud Similarity Measure (CSM) was used by the CNN model to detect faulty knowledge, resulting in improved classifier performance and supporting aircraft maintenance. By combining structured and unstructured patterns for fault diagnosis, Xu Z et al [20] were able to identify the underlying cause of the aircraft fault.

The proposed framework utilized Word2vec to extract the contents of search results and suggest results based on the similarity of their vectors. This approach allows users to view results relevant to the keywords they have entered. Similar techniques have been employed by many search engines, but with various optimization search functions.

Reddy et al [22] utilized NLP to extract non-polysemous information from a vast medical corpus, while with the proposed methodology, NLP was used to select relevant information from BBC news. Pawade [23] highlighted the importance of search engine optimization using an inverted index on a large dataset. The proposed methodology also employs the inverted index technique, but with fewer entries than Pawade [23]. However, the proposed technique supports the idea that the search function can be scaled to search big data efficiently.

The study contends that a large dataset may contain spurious correlations as a result of its size. These correlations can overpower the identification of pertinent information that cannot be differentiated algorithmically. The more extensive the features examined, the higher the likelihood that spurious correlations will result in inaccurate conclusions and dominate the ultimate outcomes.

Dilrukshi et al [26] developed a news classification and categorization system based on stratifying news by relevance. The study utilized a Twitter dataset and focused on articles related to Sri Lanka, utilizing dimension reduction techniques. Frequent words were found to be less informative for text categorization, so irrelevant words were removed as noisy data. The models were evaluated using precision and recall values for each category, and the evaluation was independently conducted for each category.

Bun and Ishizuka [27] conducted a study on news articles to group them based on their topics. They utilized the Term Frequency Proportional Document Frequency (TF-PDF) approach, which calculated the relevance of a word in a specific text. The study revealed that the significance of a word proportionally increases based on its frequency of occurrence in the text.

Kapusta and Obonya [28] conducted research on identifying fake news and categorizing news articles as either authentic or fake. Their objective was to develop a feature set from floating languages, such as Slovak, and apply it to detect fake news. They created a dataset using news articles from various publishers and labeled the features after scrutinizing the authenticity of the information. The study emphasized morphological learning approaches over contextual learning approaches. They introduced a technique to classify floating languages using Part-of-Speech (POS) tagging, which yielded reasonable accuracy.

Zhu, Y et al [29] performed sentiment analysis on Chinese comments and utilized aspect-based sentiment analysis (ABSA) for classification. They introduced a hybrid attention-based aspect-level recurrent convolutional neural network (AARCNN) model for ABSA. The proposed model was based on a sentence attention-based framework that focused on extracting relevant information from sentences.

Yousef and Voskergian [30] introduced TextNetTopics, a feature selection model that employs Bag-of-topics (BoT) instead of the BoW technique. The BoT method selects topics instead of individual words. The model utilized neural network layers to capture and analyze relevant information from sentences. The experiments demonstrated that TextNetTopics outperformed other feature selection models and achieved better results on various textual datasets.

Yaxin Bi [9] proposed a class-indifferent approach for merging classifiers outputs using evidential structures (triplet and quartet) with Dempster's rule of combination. This ensemble methodology aimed to distinguish relevant observations from irrelevant ones by representing classifier results and providing pragmatic ways to apply Dempster–Shafer theory of evidence to the ensemble learning scheme. To combine the mass functions, a formalism modeling classifier outputs as triplet mass functions was designed to provide decision support. In addition, a comparative analysis was conducted with dichotomous structures to compare the proposed method with majority voting and Dempster's rule.

The experiment was conducted using the UCI dataset, which demonstrated the advantages of the proposed approach. Table 1 presents a literature-based comparative analysis to validate the superior performance of the proposed model. According to Table 1, recent studies have not extensively explored news classification/categorization based on sentiment analysis. While many studies have utilized benchmarked datasets to discover news categories, these datasets may not be applicable in real-world scenarios. Furthermore, while much attention has been given to classifying news as authentic or fake, very few studies have focused on news classification and categorization using sentiment analysis. The proposed function, as demonstrated in Table 1, outperforms most of the papers discussed in this section. In summary, the literature review highlights these key findings.

- There is no single sentiment analysis method that performs better than others across different datasets. Additionally, there are only a few studies that explicitly apply established fusion techniques to combine classifier results for sentiment classification. Therefore, more extensive experimental work is needed to apply evidential reasoning approaches to the combination of classifiers for sentiment classification [9, 10].

- To determine sentiment classification, it is necessary to address the uncertainty that arises due to the ambiguity between positive, negative, and neutral categories. Moreover, there is a current lack of effective methods to deal with this uncertainty, particularly in the context of sentiment classification.

**TABLE 1** Literature based comparative analysis

| Author | Method | Data | Recall | Precision | Limitation |
|---|---|---|---|---|---|
| Jalil et al [21] | Sentiment analysis | COVID-19 | 93% | 95% | False negative |
| Reddy et al [22] | NLP | Medicine | 90% | 85% | Symbolic |
| Pawade [23] | Search | Internet | 76% | 63% | Polysemy |
| Huang et al [25] | Inverted index | COVID-19 | 64% | 70% | Weighting |
| Dilrukshi et al [26] | Naive Bayes | Twitter | 90% | 62% | Ambiguous |
| Bun and Ishizuka [27] | TF-PDF | News archive | - | - | Stop words |
| Kapusta and Obonya [28] | Decision Tree | Slovak text | 68% | 75% | Noise |
| Zhu, Y et al [29] | Sentiment analysis | Chinese text | 90% | 68% | Feature engineering |
| Yousef and Voskergian [30] | TextNetTopics | Textual | 80% | 71% | Misclassification |
| Yaxin Bi [11] | Triplet belief functions | Review | - | - | Not self-adaptable |

## 3 | RESEARCH METHODOLOGY

A search function was developed to retrieve BBC news information based on specific keywords (Figure 2), aiming to address the problem of polysemy and enhance search function accuracy. The web mining method was used to collect data by extracting content from web documents. A web crawler was employed to crawl the BBC news website and retrieve the required data [13, 31]. The collected data was subjected to NLTK pre-processing and inverted index phases before being recorded in the database. To improve search accuracy, a search function based on a search engine was developed to crawl the BBC news website for specific keywords. When these keywords matched the database features, corresponding results were produced. Otherwise, the most similar results were provided. The NLTK was chosen to implement the search function by tokenizing sentences into words. The inverted index helped in collecting relevant information, while Word2vector computed news similarity. The BM25, a probabilistic retrieval model developed by Stephen E. Robertson in the 1970s and 1980s, was used in the optimization scheme [32]. The BM25 method is utilized to tokenize the user's keywords into distinct words and then apply a ranking function to arrange matching information based on their significance. This method utilizes a probabilistic retrieval approach to match patterns with their corresponding indexed information [32].

To determine the similarity of each document, a score is generally calculated. The primary purpose of BM25 is to rank web documents based on specific queries, and it can also be seen as a measure of relevant information in some cases [33]. In the preprocessing phase, BM25 was implemented as a ranking method for the search engine to estimate a document's relevance for a given query. The BM25 has been found to be more adaptable compared to the traditional Term Frequency Inverse Document Frequency (TF-IDF) method [23, 24]. This adaptability of BM25 makes it more flexible.

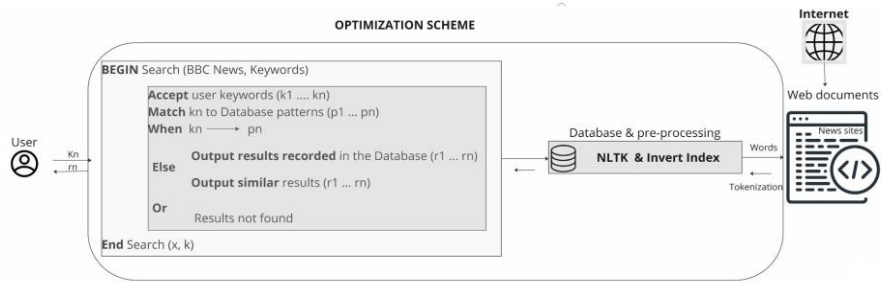

**FIGURE 2** The search function optimization scheme

## 3.1 The classification complexity of the BM25

The classification complexity of the BM25 technique can be demonstrated by analyzing its time complexity in terms of the number of documents and the length of the documents. Let N be the number of documents, *n be* the average length of the documents, and m be the number of keywords. The BM25 technique involves the following steps:

- Preprocessing: The documents are preprocessed to tokenize and stem the words, remove stop words, and calculate the term frequency.

- Query Processing: The query is also preprocessed, and the query terms are ranked based on their relevance to the query.

- Scoring: The documents are scored based on their relevance to the query.

Where $D$ is a document, $Q$ is a query, $k^i$ is a keyword, $f^{i,j}$ is the frequency of $k^i$ in document $j$, $k^{i,j}$ is a parameter that determines the saturation point of the score function for $k_i$ in document $j$, $b$ is a parameter that determines the importance of the document length.

- Ranking: The documents are ranked based on their scores, and the top-k documents are selected as the search results. The time complexity of each step can be analyzed as follows:

**Preprocessing**: The preprocessing step has a time complexity of $(Nn)$, as each document needs to be processed individually.

**Query Processing:** The query processing step has a time complexity of $O(m \log N)$, as the query terms need to be ranked based on their relevance to the query.

**Scoring**: The scoring step has a time complexity of $O(Nm)$, as each document needs to be scored individually for each keyword.

**Ranking**: The ranking step has a time complexity of $O(N \log N)$, as the documents need to be sorted based on their scores.

Therefore, the overall time complexity of the BM25 technique can be expressed as:

$$O(Nn + m\log N + Nm + N\log N$$

In practice, the value of m can be much larger than the value of n or N, which means that the time complexity of the scoring step dominates the overall time complexity of the algorithm. As a result, the BM25 technique can be computationally expensive for large datasets and queries with many keywords.

## 3.2 | The experimental data

The study selected BBC data for crawling because it is the world's largest news broadcaster[16], which lends credibility and authority to the news examined in this study [16, 34]. The BBC news website encompasses a wide range of topics that are appropriate for the proposed search function. The study randomly selected a sample of 800 articles from the website to ensure the reliability of the search results. The data were stored in a database table called news (Figure 33). The study retrieved several attributes from multiple BBC articles, including the Uniform Resource Locator (URL), content, date, title, and label, which displays the categorized information of BBC news (Figure 3). Afterward, sentiment analysis computation was performed on the database table's content attribute to determine the polarity of textual data. This method proved to be useful in predicting the sentimentality strength of BBC news.

**FIGURE 3** The news table of the database

Figure 4 displays the stop words type that was detected and ignored during the feature engineering process. The frequency of stop words occurrence in the BBC dataset is shown in Figure 5. Removing stop words during feature engineering reduced both the size of the original dataset and the time required for sentiment analysis prediction [35, 36]. The elimination of stop words enhanced the prediction accuracy by retaining only relevant tokens or patterns in the BBC dataset. This could improve the values of evaluation metrics. The dataset contains three categorical variables or columns (Covid, Vaccine, and Travel). The extracted features' descriptions (desc) using the proposed search function are presented in Figure 6.

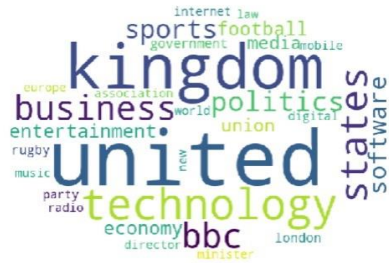

**FIGURE 4** The types of words ignored in the engineering process

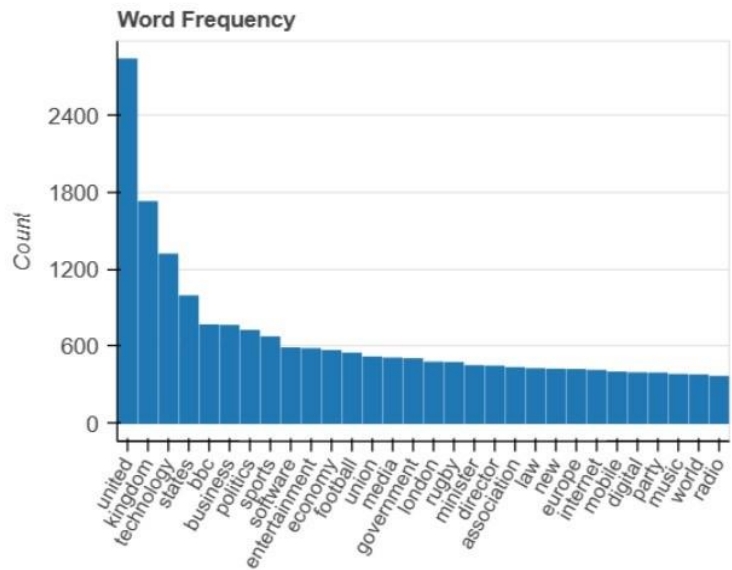

**FIGURE 5** Stop words frequency

The dataset contains 2,128 observations, with 88.3% of them being unique patterns and no missing values. The search function identified 31,406 words, with only two stop words (said and mr). The extracted features and their descriptions using the proposed search function are shown in Figure 7. In contrast, the normal search resulted in 2,410 observations with 1,709 dash punctuation and 2,150 decimal numbers, indicating the presence of noisy features. The normal search extracted a total of 13,357 words, significantly less than the proposed search function. The number of words extracted by the normal search is lower compared to the proposed search due to the difference in feature engineering. However, both datasets are balanced with a difference of only 282 observations (2,410-2,128). A comprehensive description of the characteristics of the BBC dataset has been provided in Table 2.

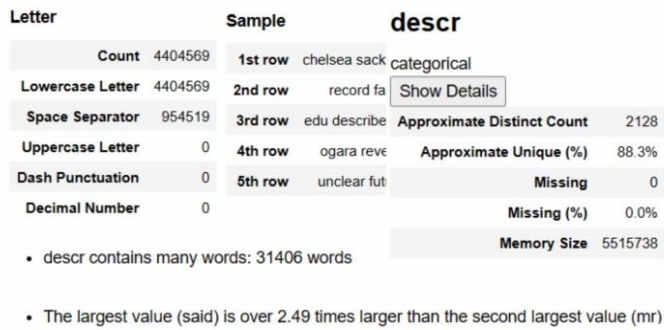

**FIGURE 6** Data extracted with the proposed search function

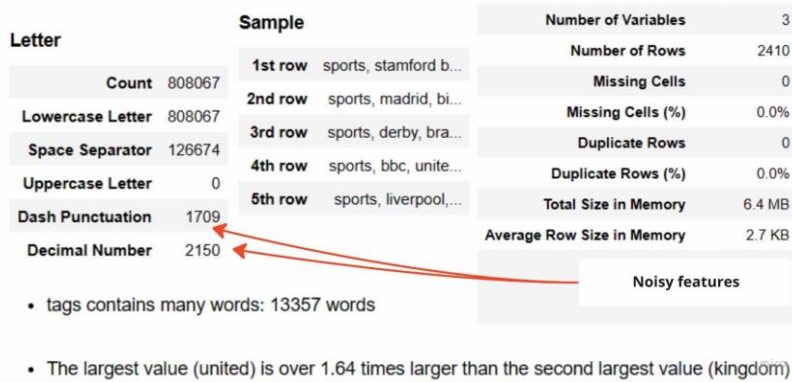

**FIGURE 7** Data extracted with a normal search function

**TABLE 2** The dataset characteristics

| No | Feature | Data type | Example | Category |
|----|---------|-----------|---------|----------|
| 1 | URL | Categorical | https://... | Covid |
| 2 | Content | Categorical | News (text) | Vaccine |
| 3 | Date | Numeric | 2023 | Travel |
| 4 | Title | Categorical | Corona minute | Covid |
| 5 | Index | Numeric | 800 | Vaccine |
| 6 | Keyword | Categorical | BBC Corona | Covid |

### 3.3 | Crawling and data cleaning

A script or program that automatically searches for information from the web based on certain rules or conditions is called a web crawler. Web crawling is advantageous for downloading web pages and achieving multi-threading. This study utilized a web crawler and specifically, the Python-based Google App Engine (PGAE) introduced by Dominic [37]. The configuration container for the web crawler included a core file that specified the interval time between two fetches, the database path directory, and the starting and ending time of crawling. The crawling frequency from the website is controlled by the interval time between two fetches. If the frequency is set too high, it may lead to the IP (Internet Protocol) address being blocked. In the data engineering process, the structure of the documents was determined using HTML div and span tags to make them identifiable. Other HTML components such as the paragraph (p) and emphasis (em) tags accurately represented the semantics of the web content. However, the use of div and span tags made the web content more accessible. The crawled features were cleaned by removing any div tags from the web pages, and the cleaned data were stored in a JSON file before being transferred to a database.

## 3.4  Feature weighting with NLTK and inverted indexing

The manual term weighting method was utilized, as Salton [38] demonstrated that manually assigning term weights is as effective as automatic weighting methods [39]. To determine the weight of a term based on its relative frequency, the proposed approach applied the Zipf law [40]. The term weighting was computed using the following formula:

$$w = log_{2(N)} - log_{2(n)} + 1$$

where N represents the total number of documents, and n is the total number of documents containing a particular term or keyword. The NLTK library in Python was used in this study with two methods, namely, NLTK tokenize and NLTK stem. The NLTK tokenize was used to segment sentences into words, using spaces to split sentences into different words.

Similarly, the NLTK stem was used to normalize words by providing them an acceptable format, such as changing the past tense to the present tense. The proposed search function tokenized the search information that users entered in the search engine using the NLTK tool, and the search was performed based on the tokenized words. Irrelevant words, known as stop words, were removed during the data cleaning phase. Stop words are words that are considered to be of little significance and are thus ignored during text analysis. The most found stop words include six determiners (a, that, the, an, and, those) which are used to describe nouns and express concepts related to localization or numbers in the text. However, removing stop words can improve computation efficiency and retrieval performance by reducing the number of indexes in the corpus. In this study, stop words were removed to improve the accuracy of the search function.

For instance, if a user searches for apples, a search engine might return 100 results. But if the user searches for bananas and apples, the search function may tokenize the sentence into three parts [banana] (and) (apples)], resulting in more than 100 results. The Python library used in this study contains pre-trained models and corpora that were utilized in implementing the search function. Inverted indexing was necessary for retrieving information from the database through indexes. Each entry in the news table includes a specific index that determines the location of textual data. An inverted index file was created beforehand, which allowed for efficient searching of information by the user.

The BBC articles were tokenized using NLTK and cleaned before being linked to a specific index and stored in the database for sentiment analysis. The index facilitates rapid search results retrieval when a user searches for certain words. The table shown in Figure 8 contains the indexes, where the keywords are recorded in the $term$ column, the number of indexes matching the keywords in the $df$ column, and the weighted index matches in the $docs$ column. For example, a user searching for BBC will receive only 697 indexes and articles as shown in Figure 8. This inverted indexing technique improves concurrency and assists in automatically generating attribute values to determine the record location.

| term | df | docs |
|---|---|---|
| sarah | 7 | [11, 1, 10][12, 1, 9][21, 1, 10][34, 1, 9][36, 1, 10][39, 1, 9][42, 1, 8] |
| evererd | 7 | [11, 1, 10][12, 1, 9][21, 1, 10][34, 1, 9][36, 1, 10][39, 1, 9][42, 1, 8] |
| disappearance | 2 | [11, 1, 10][34, 1, 9] |
| met | 7 | [11, 1, 10][329, 1, 11][338, 1, 11][443, 1, 12][502, 1, 10][628, 1, 8][702, 1, 13] |
| officer | 10 | [11, 1, 10][18, 1, 12][293, 1, 8][342, 1, 11][347, 1, 10][369, 1, 10][382, 1, 12][59 |
| arrested | 5 | [11, 1, 10][315, 1, 10][368, 1, 8][599, 1, 10][632, 1, 10] |
| suspicion | 1 | [11, 1, 10] |
| murder | 15 | [11, 1, 10][29, 1, 7][99, 1, 9][114, 1, 11][135, 1, 9][235, 1, 11][304, 1, |
| bbc | 697 | [11, 1, 10][12, 1, 9][13, 1, 9][14, 1, 8][15, 1, 8][16, 1, 9][17, 1, 9][18, 1, 12][19, 1, |
| news | 697 | [11, 1, 10][12, 1, 9][13, 1, 9][14, 1, 8][15, 1, 8][16, 1, 9][17, 1, 9][18, 1, 12][19, 1, |
| human | 1 | [12, 1, 9] |
| remains | 1 | [12, 1, 9] |
| found | 11 | [12, 1, 9][53, 1, 11][93, 1, 9][157, 1, 8][386, 1, 7][391, 1, 11][412, 1, 10][468, 1, 8 |
| kent | 3 | [12, 1, 9][34, 1, 9][99, 1, 9] |
| woodland | 1 | [12, 1, 9] |
| covid | 100 | [13, 1, 9][19, 1, 9][22, 1, 8][32, 1, 9][47, 1, 7][58, 1, 8][61, 1, 10][72, 1, 10][78, 1, |
| scotland | 17 | [13, 1, 9][22, 1, 8][32, 1, 9][121, 1, 9][153, 1, 10][161, 1, 8][192, 1, 8][367, 1, 9][ |
| rules | 6 | [13, 1, 9][19, 1, 9][71, 1, 9][123, 1, 9][321, 1, 11][669, 1, 12] |
| people | 11 | [13, 1, 9][134, 1, 11][245, 1, 9][346, 1, 8][363, 1, 9][370, 1, 7][484, 1, 10][510, 1, |
| meeting | 1 | [13, 1, 9] |
| outdoors | 1 | [13, 1, 9] |
| eased | 2 | [13, 1, 9][321, 1, 11] |
| pmqs | 3 | [14, 1, 8][428, 1, 9][661, 1, 9] |
| happened | 23 | [14, 1, 8][61, 1, 10][98, 1, 6][155, 1, 8][168, 1, 10][202, 1, 8][246, 1, 8][270, 1, 1( |

**FIGURE 8** The index table with weight

## 3.5 | Search function with sentiment analysis

The BBC news website was crawled using a web crawler to extract patterns such as date, title, and content, and the extracted news features were then stored in a local database as shown in Figure 9. In the data pre-processing phase, the news data was pre-processed, and relevant keywords were matched using the NLP module of the framework, which includes the NLTK and inverted index. The NLTK module was utilized to tokenize the BBC news into words, remove punctuation and full stops using Regular Expressions, convert all words to lowercase, and remove irrelevant words with no meaning. Additionally, stemming was used to normalize words to their primitive form, which reduces incomplete search results. The inverted index was then used to efficiently and accurately search for keywords and point to relevant BBC articles.

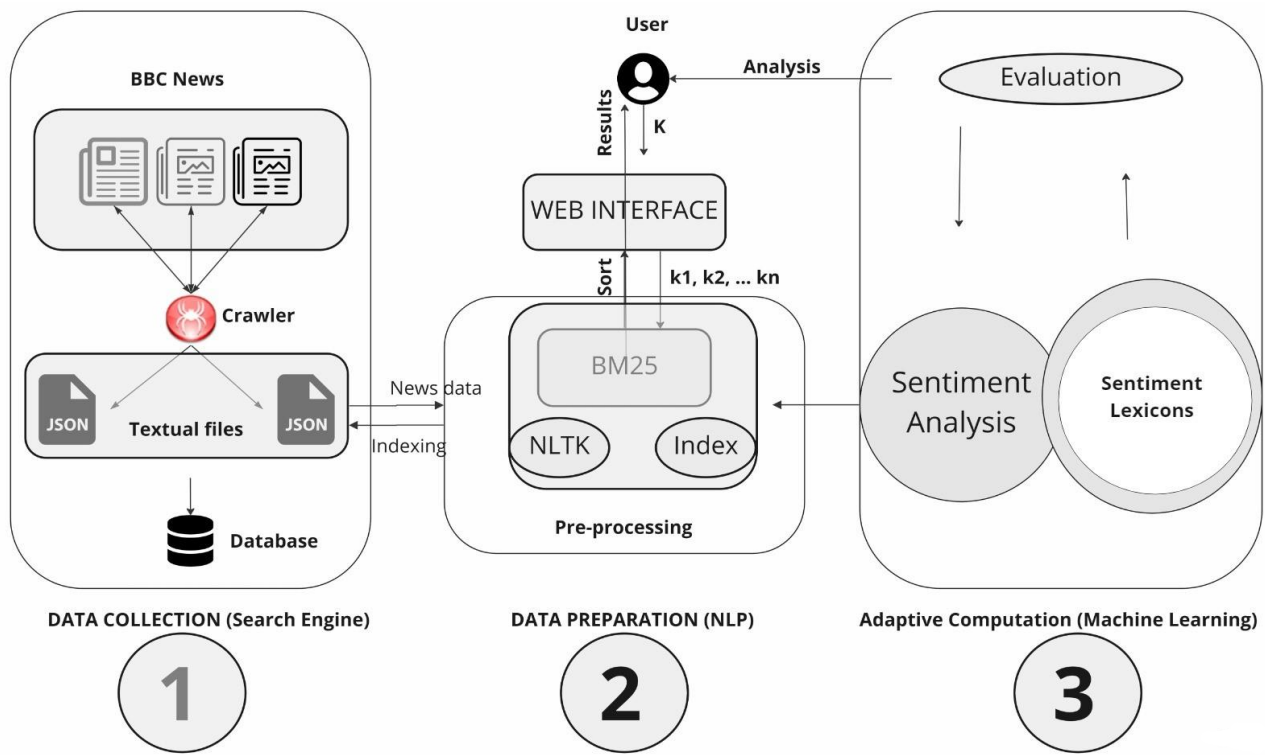

**FIGURE 9** Search function and sentiment analysis

The data wrangling phase of the framework involves a search function that is activated when keywords are entered, and the BM25 algorithm is employed to extract matching indexes from the JSON file. Each search result is assigned a score by the BM25 algorithm, which is used to measure the similarity between the keywords and the web articles. The search results are then ranked based on their scores. In the search function module, the user's keywords are compared with the pre-stored words in the index table. If there is a match, the corresponding index is retrieved, and the matching news is located. If the entered keywords do not match any of the indexed words, the most similar index and words are located using Regular Expression [41]. The interface module acts as the front-end of the framework, and HTML is used to search for content pages. When users enter their keywords, the search results are displayed on the content page.

Figure 9 illustrates the process of the search function that utilizes sentiment analysis. In the data collection phase, a web crawler is utilized to retrieve news from the BBC website, which is then stored in a JSON file. This file is subsequently transferred to the database. The pre-processing module is automated to enable the storage of new keywords in the database if they are like the embedded categories of the search function. The news features from the website are tokenized into words using the NLTK, which removes punctuation with a Regular Expression and converts words to lowercase. Sentence tokenization is applied to remove stop words.

The process of the search function using sentiment analysis is illustrated in Figure 9. The first step involves collecting data from the BBC website through web crawling and storing it in a JSON file, which is then transferred to the database. The preprocessing module is automated, such that when users enter new keywords, the system stores them in the database if they match the existing categories of the search function. The news features are tokenized into words using the NLTK and cleaned by removing punctuation and converting them to lowercase.

Stop words are removed using sentence tokenization. The text is then normalized using stemming, and cleaned words are transformed into indexes that correspond to the matched BBC article. The generated indexes are stored in a posting table in the database. In the search process, users input keywords into the web interface, which are passed to the database to search the index table for matching keywords. The system then matches the keywords found in the database with specific web articles to provide search results, which are sorted based on relevance. Finally, the sorted search results are displayed on the web page and presented to users. Overall, the search function allows users to enter a set of keywords into the web interface, which are stored in the database and evaluated for polarity using sentiment analysis.

## 3.6 │ Sentiment lexicons

A sentiment lexicon is a collection of words or phrases that are associated with specific emotions or sentiments and is commonly used in sentiment analysis algorithms. The lexicon enables the algorithm to compare input words with previously labeled words in the lexicon to predict the sentiment or polarity of sentences. One such algorithm used in the proposed framework is VADER (Valence Aware Dictionary and Sentiment Reasoner) [42], SentiWordNet [43], Sentistrength [44], Li and Hu lexicon [45], and AFINN11 [46]. The sentiment analysis models used in the study had their own set of lexicons. The framework applied each model to the BBC data and utilized their respective lexicons to match the keywords in the news articles to their corresponding polarity labels, in order to predict the overall sentiment of the BBC sentences.

## 3.7 │ Sentiment analysis algorithms

This section presents the experimental sentiment analysis algorithms and the evaluation metrics used to assess the sentiment analysis computation. The search methodology comprises two types of searches: the proposed optimized search function and the simple normal search, both of which produced a dataset. To address the class imbalance in the datasets, the study used feature engineering. The sentiment analysis on these datasets was then evaluated. The dataset was divided into training and testing samples, with 70% of the entire set assigned to data-70 and the remaining 30% to data-30, which served as a validation set. This approach ensured that the sentiment analysis model was not trained using the testing sets and that the validation results were independent of any discrepancies and biases. The study compared the proposed search function to the normal search procedure based on the data collected from the web. The evaluation metrics used to assess the overall experiments were precision, accuracy, recall, and F1 score. Figure 10 depicts the differences between the two searches.

|  | Embedded category | Words category | Methods |
|---|---|---|---|
| Classify search function | Yes | Covid, vaccine and travel | BM25, NLTK, inverted index, and sentiment analysis |
| Normal search function | No | | Sentiment analysis |

**FIGURE 10** The type of searches

The primary difference between the two search methods is the inclusion of embedded categorization. The same set of keywords, such as Covid, Vaccine, and Travel, are utilized for both search methods to ensure result reliability. The search process is divided into three clusters: Covid, Vaccine, and Travel. The Covid category contains coronavirus news stored using the Covid keywords (Figure 11). The Vaccine category includes Covid vaccine news stored using the vaccine keywords (Figure 11).

Lastly, the proposed search function stores Covid-related travel news using the Travel keyword in the Travel category (Figure 12). The computing environment used in this study is presented in Table 3. The evaluation of the two search functions is based on the data collected from the web, and metrics such as precision, accuracy, recall, and F1 score are used to assess the overall

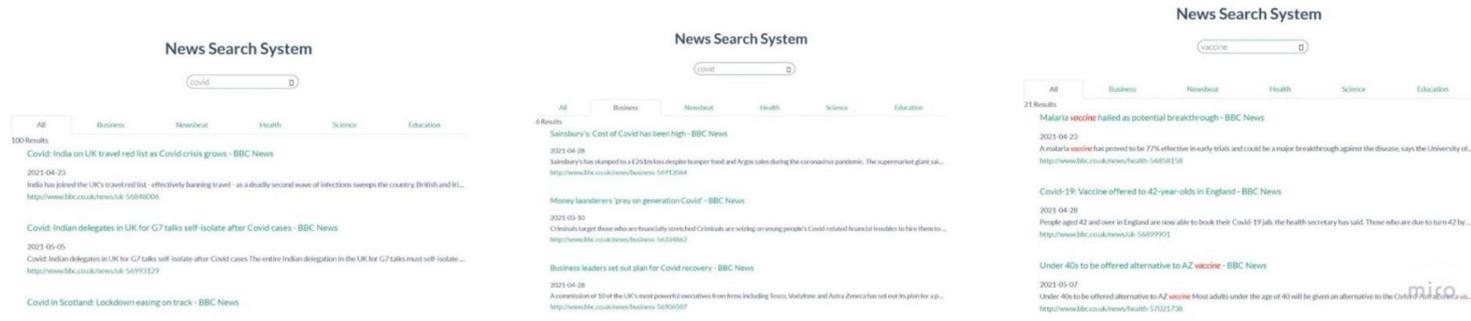

experiments.

**FIGURE 11** The searches results and data collected

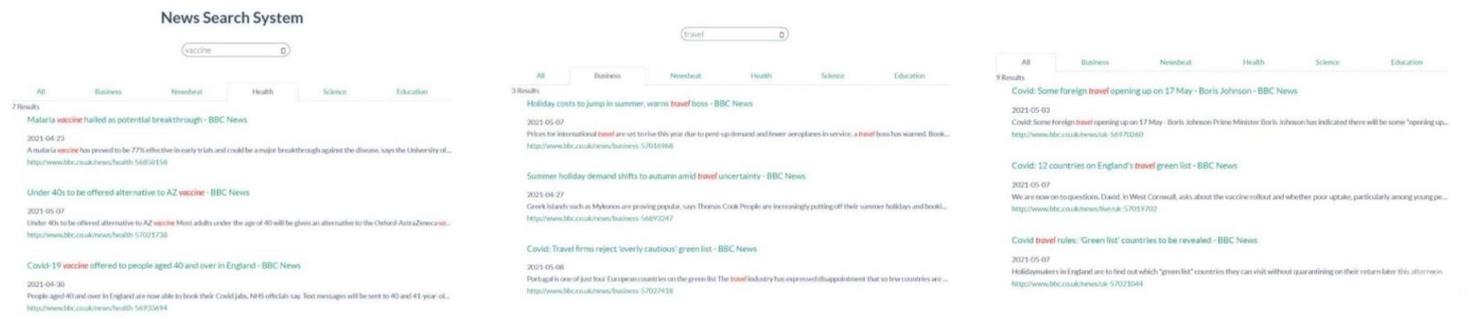

**FIGURE 12** The data collected using keywords

**TABLE 3** The computing environment

| | Component | Type | Example |
|---|---|---|---|
| 1 | Search engine | PGAE-based BBC website | Proposed search |
| | | BBC web site | Normal search |
| 2 | Operating system | Windows & Linux | Windows 10 & Debian |
| 3 | Algorithm | VADER, SentiWordNet, Sentistrength, Li & Hu, and AFINN-11 | Classification |
| 4 | Sentiment analysis | Sentiment lexicons | Training |
| 5 | NLP | Inverted index, word2vec, NLTK, & BM25 | Text segmentation & search |
| 6 | Cross-validation | 70% (training) & 30% (testing) | K-fold |
| 7 | Data | BBC news | Corona text news |
| 8 | Language | Python | Python 3.9.0 |

## 3.8 | VADER

VADER is a sentiment analysis program that uses a rule-based lexicon designed specifically for social media analysis. Its developers, Hutto and Gilbert [47], created the algorithm to address the challenges in analyzing symbols, languages, and text styles that are commonly found in social media. VADER is capable of detecting positive, negative, and neutral polarities of textual data. The VADER lexicon is built by selecting textual patterns from the General Inquirer (GI) and Linguistic Inquiry and Word Count (LIWC) preset lexicons, and includes slang, social media abbreviations, emojis, and facial expressions such as :-) to represent positive, negative, or neutral sentiments. The lexicon consists of 7,500 textual features. To compute the polarity of a sentence using VADER, a compound score is calculated by summing equivalent values for each word in the lexicon and adjusting them according to specific rules. The polarity score is then restricted to be between the most negative (-1) and most positive (+1) values. The assigned polarity values for each row in the analyzed text are positive (pos), negative (neg), and neutral (neu), as shown in Figure 13. VADER uses a set of thresholds to assign positive, negative, or neutral polarity, as shown in Figure 14.

```
#from vaderSentiment import SentimentIntensityAnalyzer
sentences=
["The book was good" #positive example
,"VADER is not smart nor funny." #negative example
,"Today only kinda sux! But I'll get by, lol" #mixed
example ]

analyzer = SentimentIntensityAnalyzer()
for sentence in sentences:
scores = analyzer.polarity_scores(sentence)
print(sentence + str(scores))

The following table shows the result after applying the previous code:

-----------------------------------------------------------
Sentence                            pos      compound   neu      Neg
-----------------------------------------------------------
The book was good                   0.492    0.4404     0.508    0.0
VADER is not smart nor funny.       0.0      -0.7424    0.354    0.646
Today only kinda sux!But I'll get by, lol  0.317  0.5249  0.556   0.127
-----------------------------------------------------------
```

**FIGURE 13** The polarity computation using VADER

```
-----------------------------------------------------------
(i)   Positive: compound score > = 0.0.5
-----------------------------------------------------------
(ii)  Neutral: (compound > -0.05) and (compound < 0.05)
-----------------------------------------------------------
(iii) Negative: compound score <= - 0.05
-----------------------------------------------------------
```

**FIGURE 14** The polarity threshold

## 3.9 | SentiWordNet

SentiWordNet is an opinion-mining tool widely used in sentiment analysis that employs a lexicon derived from WordNet. This lexicon is divided into synonyms (synsets), nouns, verbs, adjectives, and other grammatical categories. The SentiWordNet algorithm utilizes the WordNet synset dictionary to combine polarity scores and determine the sentiment of the text as either negative, objective (neutral), or positive.

The algorithm generates three scores, each ranging from zero to one, using supervised machine learning methods [43]. The PosScore and NegScore represent the level of positivity and negativity associated with the text. The process depicted in Figure 15 has been used to assign sentimentality using SentiWordNet version 3.0, as explained in Figure 15. To determine whether a given text feature has a positive, negative, or neutral sentiment, the average scores of its associated synsets were used. If the positivity average score exceeds the negativity average score, then the sentiment is considered positive.

The lexicon used contained 64,000 features, and a Python code was written utilizing the NLTK package. The calculation of sentimentality for expressions in the lexicon is shown in Figure 16.

```
----------------------------------------------------
PosScore [0,1] : positivity scale
NegScore [0,1]: negative scale
ObjScore [0,1]: objective/neutral scale
----------------------------------------------------

Where the degree of objectivity can be calculated as
follows:

ObjScore = 1 - (PosScore + NegScore).
----------------------------------------------------
```

**FIGURE 15** The SentiWordNet computation

```
----------------------------------------------
<very.r.01: PosScore=0.25 NegScore=0.25>
<nice.a.01: PosScore=0.875 NegScore=0.0>
<love.v.02: PosScore=1.0 NegScore=0.0>
<worse.a.01: PosScore=0.0 NegScore=0.75>
<bad.a.01: PosScore=0.0 NegScore=0.625>
<truly.r.01: PosScore=0.625 NegScore=0.0>
----------------------------------------------
```

**FIGURE 16** The polarity of SentiWordNet

## 3.10 | Sentistrength

Sentistrength is an open-source sentiment analysis tool that can detect the emotions expressed in text. It has been used in the Cyber-Emotions project to analyze over 14,000 social media posts with a level of accuracy comparable to that of a human. The Sentistrength lexicon is based on terms and expressions derived from the Linguistic Inquiry and Word Count (LIWC) [48, 44, 49]. Each textual pattern is assigned scores for negative, positive, and neutral polarities. Sentistrength predicts the negative and positive polarities in short texts using specific rules, which are:

- Negative (-1) to extremely negative (-5)

- Positive (1) to extremely positive (5)

The Sentistrength algorithm requires placing textual features in a plain text file (one text per line) to predict text polarities. The resulting output is a copy of the file containing negative and positive predictions at the end of each line. As shown in Figure 17, the negative and positive values of each text can be predicted using Sentistrength. The final and averaged polarity of the file can be obtained with different point scales ranging from -5 (negativity) to +5 (positivity), where 0 represents neutral polarity.

Previous experiments have demonstrated the robust performance of Sentistrength in web mining [44, 49]. This tool is available in both Java and Windows formats, with this study utilizing the publicly accessible Windows version, which can be found at http://sentistrength.wlv.ac.uk/. The Sentistrength lexicon is presented in Figure 18.

| Text | Positive | Negative |
|------|----------|----------|
| using Linux and loving it - so much nicer than windows... Looking forward to using the wysiwyg latex editor! | 4 | -1 |
| @ruby_gem My primary debit card is Visa Electron. | 1 | -1 |
| @kirstiealley I hate going to the dentist.. !!! | 1 | -4 |

**FIGURE 17** The Sentistrength computation

| List Name | Sample Word | Score |
|-----------|-------------|-------|
| Sentiment strength word list | Awful | -4 |
| | Blissful | +5 |
| Booster word list | Slightly | -1 |
| | Extremely | +2 |
| Idiom List | Shock horror | -2 |
| | Whats good | +2 |
| Negation word list | Cant | - |
| | Never | - |
| Emoticon word list | :'( | -1 |
| | :-D | +1 |

**FIGURE 18** The Sentistrength lexicon

### 3.11 | Liu and Hu lexicon

The Liu and Hu lexicon used in this study consists of two-word lists: one for positive predictions and the other for negative predictions [50]. It was developed by researchers from the Computer Science Department at the University of Illinois [50]. The Python code used to access the negative and positive lexicon words is demonstrated in Figure 19. To predict the sentiment of texts, a Python code was created to compare the BBC texts with the labeled lexicon words using the NLTK.

```
>>> from nltk.corpus import opinion_lexicon
>>> opinion_lexicon.words()
['2-faced', '2-faces', 'abnormal', 'abolish', . . .]
>>> opinion_lexicon.negative()
['2-faced', '2-faces', 'abnormal', 'abolish', . . .]
>>> opinion_lexicon.positive()
['a+', 'abound', 'abounds', 'abundance', 'abundant', . . .]
```

**FIGURE 19** The Liu and Hu lexicon

### 3.12 | AFINN-111

AFINN-111 consists of a collection of English words that have been assigned a sentiment score ranging from -5 (negative) to +5 (positive) [51]. This lexicon was developed manually by Finn Arup Nielsen over the course of 11 years from 2000 to 2011 [51]. The lexicon includes two lists, AFINN-111 containing 2,477 phrases and words, and AFINN-96 with 1,468 unique phrases and words consisting of 1,480 lines. The list includes two types of columns: the word itself and its corresponding polarity value, which is depicted in the range of [-5, +5]. Figure 20 displays the AFINN-111 list.

```
-----------------------------------------
Word                        Value
-----------------------------------------
Lose                        -3
Loses                       -3
Loser                       -3
Losing                      -3
Loss                        -3
Lost                        -3
Lovable                      3
Love                         3
Loved                        3
Loving                       2
Lowest                      -1
-----------------------------------------

The following Python code demonstrates how to calculate
the degree of emotion of a positive statement of degree 3:

-----------------------------------------
afinn= dict(map(lambda kv:(kv[1])),[line.split('\t')
for line in open("AFINN-111.txt")]))
score=sum(map(lambda word: afinn.get(word, 0), "
AFINN is very good".lower().split()))
print(score) // 3
-----------------------------------------
```

**FIGURE 20** The AFINN-111 lexicon and computation

Based on the analysis conducted, SentiWordNet was found to have the highest occurrence of positive and neutral textual patterns (Figure 21), whereas Sentistrength exhibited the highest frequency of negative patterns (Figure 21). It is important to note that this finding suggests that the size of sentiment analysis lexicons can significantly impact the classification or prediction results obtained, and researchers should carefully consider the size and characteristics of each lexicon before selecting. Additionally, it highlights the importance of selecting appropriate lexicons for sentiment analysis to ensure the accuracy and validity of the results. It is important to select appropriate lexicons for sentiment analysis because the accuracy and effectiveness of sentiment analysis largely depend on the quality and suitability of the lexicon used. A lexicon is a dictionary or database that contains words and phrases along with their assigned sentiment polarity (positive, negative, or neutral) [11]. Using an inappropriate lexicon can lead to inaccurate sentiment analysis results, which can have serious consequences in various domains such as business, politics, and healthcare. For example, misinterpreting customer sentiments in business can result in a loss of revenue and reputation, while misinterpreting patient sentiments in healthcare can lead to incorrect diagnoses and treatments. Therefore, selecting the right lexicon that is tailored to the specific domain and language is crucial for obtaining accurate sentiment analysis results. It is also important to continuously update and improve the lexicon as language and context evolve over time.

## 3.13 | Evaluation metrics

In the experiments, precision (P), accuracy (A), F1 score (F), and recall (R) [52, 53] are used as evaluation metrics. Precision represents the number of true predicted classes that belong to the accurate class. Recall, on the other hand, determines the percentage of correctly classified categories [52, 54]. It is the number of observations that the sentiment analysis model correctly predicted divided by the total number of observations [55, 56]. Recall measures the number of instances that the sentiment analysis model predicted correctly and are in a particular class, divided by the total number of instances that belong to that class [55, 57]. True positive (TP) refers to instances where the model accurately predicts the positive class, while true negative (TN) refers to instances where the model accurately predicts the negative class [58, 59]. False positive (FP) refers to instances where the model incorrectly predicts the positive class, and false negative (FN) refers to instances where the model incorrectly predicts the negative class [60, 61]. Accuracy measures the overall classification/prediction success of the sentiment analysis.

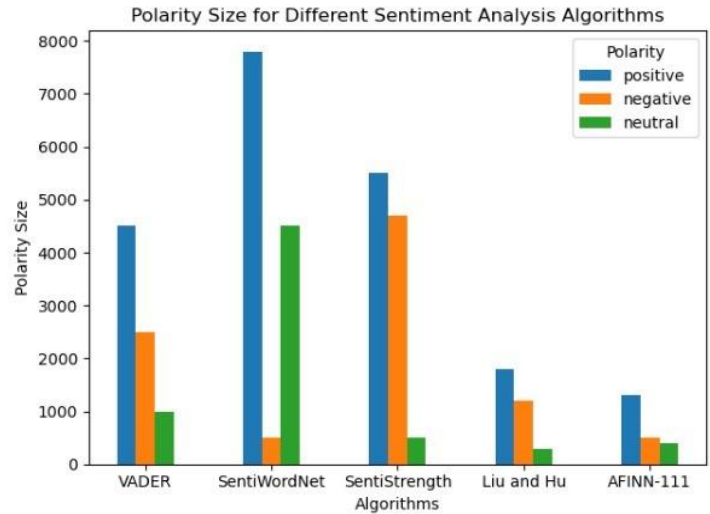

**FIGURE 21** The size of lexicons

# 4 | RESULTS

The results of sentiment analysis on the BBC data are presented in this section. The experiment employed sentiment analysis lexicons without any modifications to analyze the data. Additionally, a comparative analysis was conducted to determine the best performing sentiment analysis algorithm. The result displays the evaluation metrics obtained from both BBC samples. The sentiment analysis performance is summarized in Figure 22. The lexicons were used without any modifications to analyze the data obtained using the proposed search function. The results showed that the VADER model had the highest accuracy of 85%. The F1-score of the SentiWordNet model was 65% using the data obtained through the proposed search. The AFINN-111 model performed well in the positive sample rating, achieving an accuracy of 78% with the proposed search data. However, the Sentistrength model did not perform well in classifying positivity/neutrality with the proposed search data, with an accuracy range of 10%-15%. When analyzing data obtained through a normal search, the precision of the AFINN-111 and VADER models was higher and closer than that of the other three models. Furthermore, it is worth noting that the performance of the Liu-Hu model's lexicon improved when using data extracted with the proposed search, as compared to a normal search. Overall, the utilized sentiment analysis lexicons and models showed promising results, even without preprocessing the data extracted with a normal search. The comparative analysis suggests that the VADER lexicon is a strong candidate for accurate classification of positive and negative news. The experiment also demonstrates the AFINN-111 model's proficiency in rating more positive samples when using the proposed search data. Furthermore, the Sentistrength model achieved the highest accuracy (75%) in classifying negative samples. However, the Liu and Hu model performed poorly compared to other techniques.

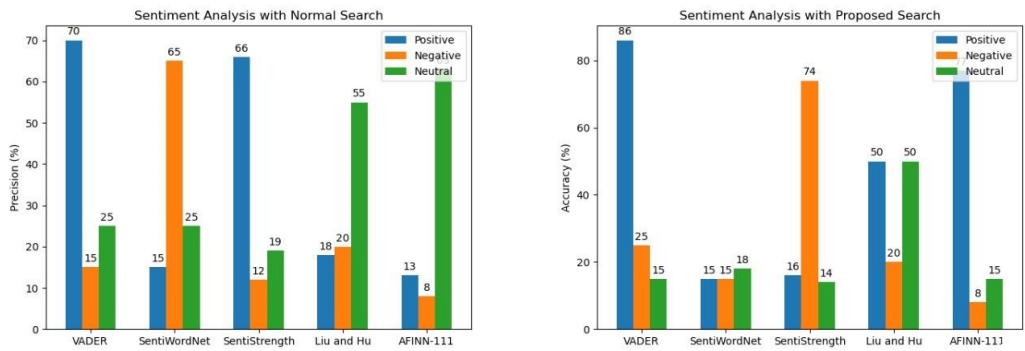

**FIGURE 22** The sentiment analysis results

The performance of the VADER model varied with the two different BBC samples, and the model's performance decreased when using data extracted with a normal search. Figure 23 illustrates that the VADER model had a misclassification rate of 15.8% for features extracted by the proposed search and a misclassification rate of 20.3% for patterns retrieved by a normal search. These results demonstrate the effectiveness of the proposed search function in enhancing the structure and quality of BBC data compared to a normal search. The NLP techniques implemented in the data preprocessing stage led to a decrease in the misclassification rate of sentiment analysis.

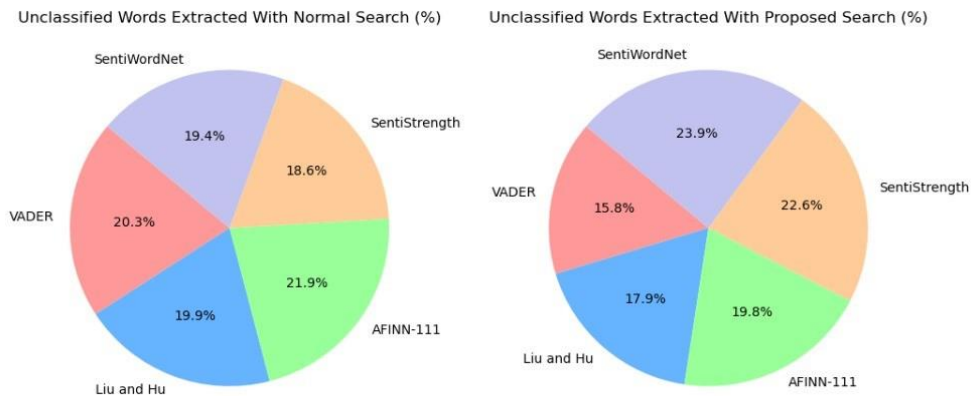

**FIGURE 23** Unclassified words

The optimal confusion matrix for sentiment analysis is depicted in Figure 24, illustrating the number of textual characteristics that have been accurately predicted/classified by the sentiment analysis as belonging to the previously mentioned news categories. Specifically, the confusion matrix for the best sentiment analysis technique, such as SentiWordNet, reveals that out of 16,744 textual features, most of them were correctly predicted to belong to the Vaccine category. SentiWordNet outperformed other models by predicting 16,744 words with a negative polarity of 68% accuracy or precision concerning the Covid vaccine. Among the sentiment analysis techniques, AFINN-111 required the longest execution time (50 seconds).

Therefore, the efficiency of sentiment analysis models in terms of time complexity and misclassification rates can serve as a crucial performance indicator for news categorization and classification. In general, the evaluation metrics exhibit a minor improvement when utilizing data obtained through the proposed search function, as illustrated in Figure 25.

The outcomes indicate that the proposed search function retrieves a greater number of URLs, content, and indexes than a standard search. This finding highlights the significance of feature engineering in news classification and categorization. By utilizing NLP for feature engineering, the accuracy and precision of sentiment analysis classification can be enhanced. Crawling with a standard search required more time compared to the proposed search, mainly due to the absence of pre-processed patterns in a standard search. The primary reason for the superior performance of the proposed search function compared to a standard search is the categorization of features recorded by the proposed search function. As a result, sentiment analysis models were able to achieve a higher level of accuracy and precision.

Among the five models, VADER implementation, SentiWordNet, and Sentistrength techniques achieved the most promising outcomes in terms of the number of URLs and textual content classified. VADER employed 8,000 textual patterns, followed by the SentiWordNet and Sentistrength models. Adjectives were found to be the most relevant linguistic pattern for sentiment analysis. Nevertheless, combining adverbs and adjectives can also enhance the sentiment analysis classification. An essential issue, however, is that adjectives and adverbs should not carry equal weight in predicting the sentiment of a given text. Rather, the scores of adjectives and adverbs should be combined using appropriate feature weighting techniques to achieve optimal results. The evaluation of five sentiment analysis models using available lexicons revealed that Sentistrength and SentiWordNet obtained comparable accuracy and precision for both types of searches, while Liu and Hu implementation achieved good accuracy and precision levels for Covid and Travel categories.

These findings led to two significant conclusions: (i) VADER-based sentiment analysis model can achieve similar accuracy for both types of BBC data; and (ii) adverbs can improve precision levels when weighted and combined with adjective scores. Adverbs were identified as the most critical linguistic feature for sentiment extraction. The use of NLP lexicons facilitated sentiment analysis computation, highlighting the importance of using a smart search function that can retrieve non-polysemic search results and provide accurate automated sentiment analysis.

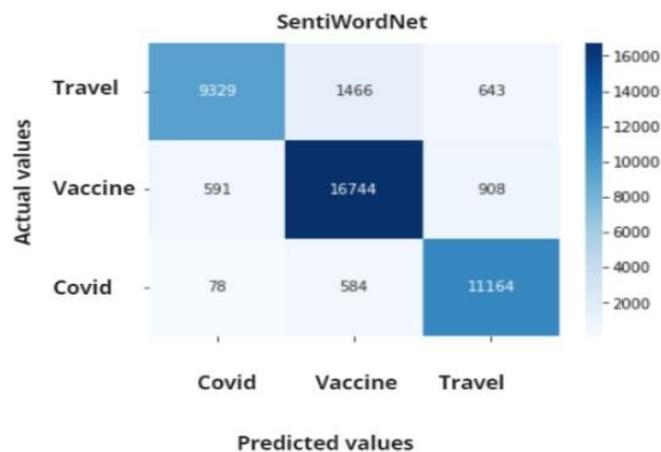

**FIGURE 24** The confusion matrix of the SentiWordNet model

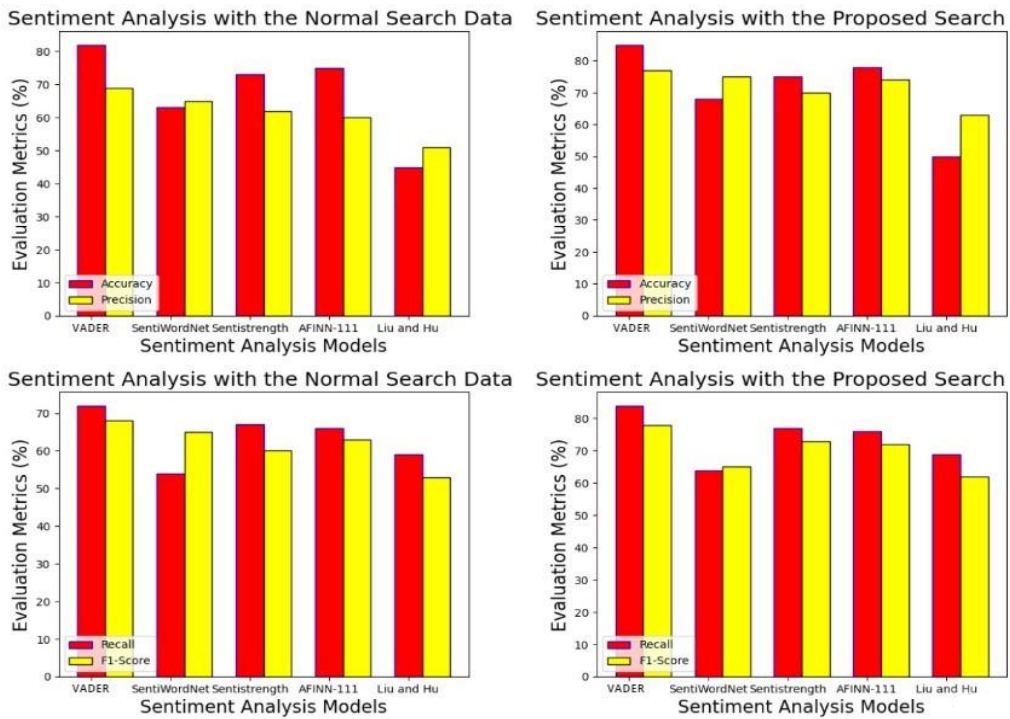

**FIGURE 25** Summary of the sentiment analysis performance

## 4.1 | Discussion

To summarize, the study found that VADER was the most efficient and accurate sentiment analysis model for news classification (Figure 26). AFINN-111 performed well in predicting positive polarity with a high accuracy of 78% when using structured data retrieved with the proposed search function (Figure 26). Liu-Hu's model did not perform well due to the limitations of its lexicons. Overall, the VADER, SentiWordNet, Sentistrength, and AFINN-111 lexicons showed promise for accurately classifying news based on their positive and negative sentiment. The best-performing model in classifying more positivity for the Covid category was the AFINN-111 with 78% accuracy (Figure 26). This result confirms the effectiveness of the proposed search function in improving the structure and quality of textual data compared to a normal search. The experiment shows the confusion matrix of the best sentiment analysis technique such as SentiWordNet which considered more textual features that were correctly predicted to belong to the Vaccine category.

The proposed search function was effective in improving the structure and quality of textual data compared to a normal search. The SentiWordNet technique had the best confusion matrix and was able to correctly predict more textual features to belong to the Vaccine category. The study suggests that the time complexity, lexicon's structure, and misclassification rates of sentiment analysis models are important indicators for gauging the efficiency of a sentiment analysis framework.

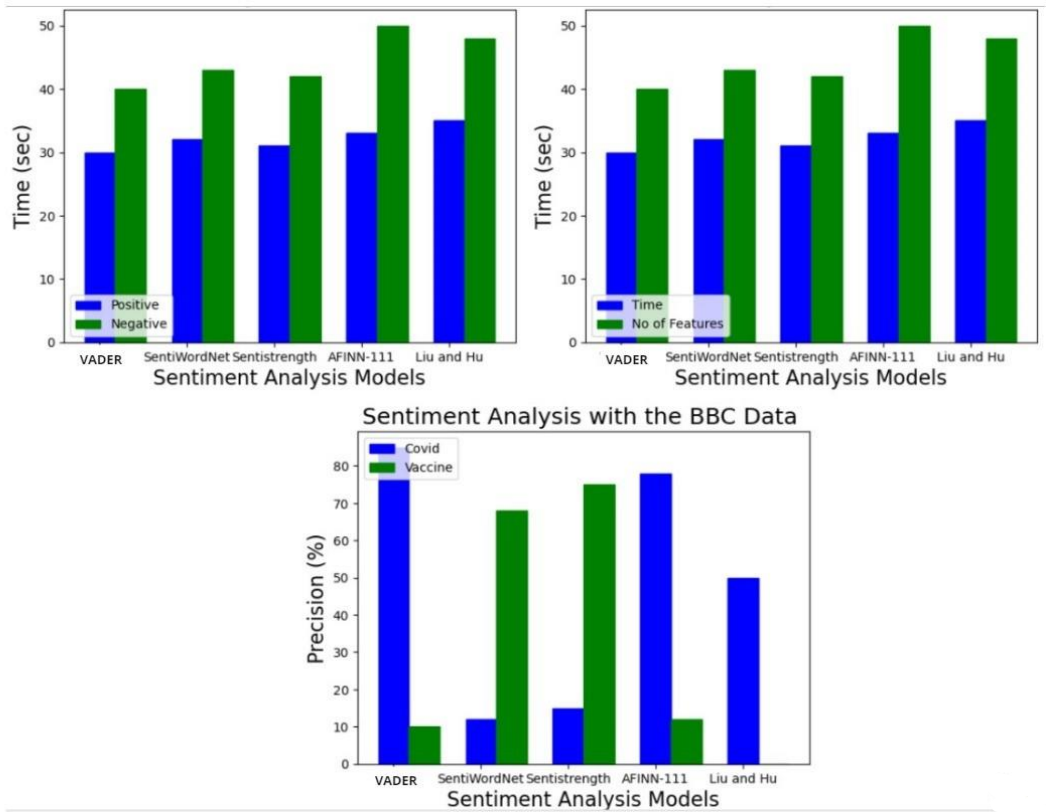

**FIGURE 26** Time execution and precision of sentiment analysis

Among the five models, VADER, SentiWordNet, and Sentistrength obtained the best results, outperforming some existing techniques. However, the study's focus was on using sentiment analysis as a subset of NLP to enhance the search engine's performance and provide users with a better perspective on the crawled data. The study used sentiment analysis in combination with other NLP techniques such as text segmentation and POS to optimize the search engine's smart search function. The aim was to develop a search engine with embedded sentiment analysis that could retrieve non-polysemic results and classify them using sentiment analysis models [43, 64, 65, 47, 66, 67]. A search engine with embedded sentiment analysis can automatically classify emotions from search results into negative, positive, or neutral categories. However, accurately determining the valence or emotions of text can be difficult due to the subjectivity of language, leading to false positive and negative rates. This means that even with updated lexicons, sentiment analysis models may still have issues predicting the polarity of statements. Sentiment analysis has various practical applications in search engine optimization. For instance, it can be used for real-time assessment of brand reputation, improving customer support interactions, identifying customer needs and anomalies, and monitoring customer behavior using web intelligence.

Nevertheless, the absence of lexicons in languages like French, Spanish, Chinese, and Swahili will impede the deployment of smart search engines and the contextualization of sentiment analysis outcomes. This will lead to inaccuracies in sentiment computation due to false positives and negatives. Moreover, building a robust and smart sentiment analysis framework for search engines requires tackling challenges such as polysemy, irony, sarcasm, and multi-polarity.

## 5 | CONCLUSION

This manuscript presents a topic that involves several aspects of NLP and sentiment analysis. The author has employed a web crawler to collect BBC news across the internet, carried out pre-processing of text by using NLP and sentiment analysis methods to determine polarity of the processed text data. The BBC website was utilized to crawl for news data using a proposed smart search function. The experimental results portray this search function collecting more than 2,000 textual patterns for sentiment analysis. The proposed function outperformed a normal search in terms of features quality. However, this study can be extended by embedding more categories in the search function to collect a large amount of data from the entire internet. In fact, the development of effective methods and functions for each aspect aforementioned in this study is always challenging. Even if experiments were performed with a single database, the author still believes that some of the results presented in this article will be used as a reference in the field of search engine optimization. Different users will have different needs, and traditional search methods cannot satisfy all needs due to polysemous problems, which lead to poor user experiences. To address this issue, the research implemented a search function enabling users to extract relevant BBC information that meets their needs. In addition, the web has a rich and diverse set of news, including fake ones. Unfortunately, the proposed search function cannot remove fake news from the search results. Another limitation is linked to the number of categories embedded in the proposed search, which do not include all groups that are relevant to users. However, the proposed list of categories such as Covid, Vaccine, and Travel can still be extended to improve the search efficiency and reliability. Furthermore, the proposed function can be improved by adding an advanced classification scheme to the search with intelligent tagging. The tagging mechanism should be automated with machine learning where the classifier can be trained to tag searched contents. The optimized search function using machine learning can also be used to study the strength of an entity's web presence. In future works, one can automatically create search categories using artificial intelligence and predict users' behavior with machine learning. An online corporate reputation analysis can also be implemented with the proposed methodology. Additional lexicons such as GPOMS (Google Play Opinion Mining Score) and Opinion Finder can also be added to the sentiment analysis models to improve the sentiment analysis performance. Lastly, ensemble learning can also be utilized to improve the machine learning accuracy on the collected BBC data.

## ACKNOWLEDGMENT

The author expresses gratitude to the University of Pretoria's Faculty of Engineering, Built Environment, and Information Technology for their support in funding this research project through the Doctorate University Capacity Development Program (UCDP) Grant A1F637. Additionally, the author acknowledges the anonymous reviewers for their invaluable suggestions and comments, which greatly contributed to enhancing the clarity and quality of the manuscript.